\documentclass[a4paper,fleqn,usenatbib,useAMS]{mnras}

\usepackage{graphicx}	
\usepackage{amsmath}	
\usepackage{amssymb}	
\usepackage{multicol}        
\usepackage{bm}		
\usepackage{pdflscape}	

\usepackage[T1]{fontenc}
\usepackage{ae,aecompl}

\usepackage{newtxtext,newtxmath}

\title[Spatial Distribution of Quasi-biennial Oscillations]{The spatial distribution of quasi-biennial oscillations in the high-latitude solar activity}

\author[L. H. Deng et al.]{
L. H. Deng,$^{1,3,5}$
Y. Fei,$^{4}$
H. Deng,$^{2}$
Y. Mei,$^{2,5}$
and F. Wang$^{1,2}$\thanks{E-mail: fengwang@gzhu.edu.cn (FW)}
\\
$^{1}$Yunnan Observatories, Chinese Academy of Sciences, Kunming~650216, P.R.~China\\
$^{2}$Center For Astrophysics, Guangzhou University, Guangzhou~510006, P.R.~China\\
$^{3}$Chongqing University of Arts and Sciences, Chongqing~402160, P.R.~China\\
$^{4}$School of Statistics and Mathematics, Yunnan University of Finance and Economics, Kunming~650221, P.R.~China\\
$^{5}$CAS Key Laboratory of Solar Activity, National Astronomical Observatories, Beijing~100012, P.R.~China}

\date{Accepted 2020 April 15. Received 2020 April 15; in original form 2020 February 04}

\pubyear{2020}

\begin{document}
\label{firstpage}
\pagerange{\pageref{firstpage}--\pageref{lastpage}}
\maketitle

\begin{abstract}

Quasi-biennial oscillations (QBOs) are considered as a fundamental mode of solar magnetic activity at low latitudes ($\leq50^\circ$). However, the evolutionary aspect and the hemispheric distribution of solar QBOs at high latitudes ($\geq60^\circ$)  are rarely studied. Here, a relatively novel time-frequency analysis technique, named the synchrosqueezed wavelet transform, is applied to extract the main components of the polar faculae in the northern and southern hemispheres for the time interval from August 1951 to December 1998. It is found as the following: (1) Apart from the 22-year Hale cycle, the 17-year extended activity cycle, and the 11-year Schwabe cycle, the QBOs have been estimated as a prominent timescale of solar magnetic activity at high latitudes; (2) the QBOs of the polar faculae are coherent in the two hemispheres, but the temporal (phase) and the spatial (amplitude) variations of solar QBOs occur unevenly on both hemispheres; and (3) for the 11-year period mode, the northern hemisphere begins three months earlier than that in the southern one. Moreover, the spatial and temporal distributions of the hemispheric QBOs differ from those of the 11-year Schwabe cycle mode in the two hemispheres. Our findings could be helpful to improve our knowledge on the physical origin of the spatial distribution of solar QBOs at high latitudes, and could also provide more constraints on solar dynamo models introduced to characterize the different components of the solar magnetic activity cycle.
\end{abstract}

\begin{keywords}
Sun: activity -- Sun: oscillations -- Sun: faculae, plages
\end{keywords}




\section{Introduction}

The Sun, a constantly variable star, exhibits perpetual changes in the level of magnetic activity manifested by dark sunspots or pores, bright faculae, solar flares, filaments or prominences, coronal mass ejections, magnetic bright points, and so forth \citep{2006RPPh...69..563S,2019MNRAS.489.3183C}. The cyclic or quasi-periodic behavior of solar magnetic activity features, which are considered to be driven by the dynamo action in the solar interior, is a very significant aspect in solar and space physics fields, but the detailed physical processes that controlled the solar activity cycle are still far from drastically understood and explained \citep{2014SSRv..186..491J, 2017ApJ...843..111C}. The statistical and observational aspects on the solar activity cycle could be retrieved from the temporal variability of various observed phenomena, such as the quasi-periodic processes of the Sun \citep{2014SSRv..186..105E,2014AJ....148...12X,2015MNRAS.451.4360K}.


Except for the nearly 11-year solar cycle \citep{1844AN.....21..233S} and the 27-day rotation cycle \citep{2013ApJ...773L...6S,2014AJ....148..101S}, the question on the exactly physical origin and the temporal evolution (or distribution) of some other periodicities is under debate. Naturally, the most important and significant oscillations with an order of around 2 years, which were named as the quasi-biennial oscillations (QBOs) of the Sun, have been gained increasing attention and interest in recent decades \citep{2004ARep...48..678B, 2006AdSpR..38..484B, 2014SSRv..186..359B}. Solar QBOs appear ubiquitous in the existing large volumes of observational data pertaining to the magnetic field \citep{1992SoPh..137..167O, 2001ARep...45.1012O, 2005SoPh..226..359C} and have been discovered in magnetic activity structures (e.g., sunspot-related and flare-related indicators) \citep{1979Natur.278..146S, 2004SoPh..222..199S, 2019RAA....19..131X,2010ApJ...709L...1V}, the solar atmosphere (mostly the photosphere and chromosphere)  \citep{1981SoPh...71..259H, 2000SoPh..197..157B, 2005SoPh..229..359V,2008JASTP..70.2112R,2011NewA...16..147G,2008AdSpR..41..297V}, and indeed the interplanetary medium, including the interplanetary magnetic field intensity, the galactic cosmic rays, and solar wind vecloticy \citep{1994GeoRL..21.1559R, 1996SoPh..167..409V, 2018AJ....156..152X, 2003JGRA..108.1367K, 2012ApJ...749..167L}. However, the oscillatory modes in the range of around 2 years have been called as the mid-term \citep{2007MNRAS.374..282F,2011SoPh..271..169M,2018Ap&SS.363...84M} or the intermediate periodicities in many articles \citep{2000AdSpR..25.1939M, 2003SoPh..212..201M, 2004SoPh..221..337M}.

Solar QBOs are directly connected with the high-frequency component of the dynamical process inside the Sun, but the exactly physical origin is still unknown. To reasonably reveal the QBOs existed in the surface magnetic field of the solar atmosphere,  \cite{1998ApJ...509L..49B} proposed a model to confirm the existence of two different types of solar dynamos running at different depths. The first one of the dynamo action is operated at the base of the convection zone, and the other one locates at the bottom of the layer extending $5\%$ (about 35000 km) below the photosphere. Furthermore, this feature can be able to explain the QBOs observed in the helioseismic data, as the temporal variation of the helioseismic data is also connected with the solar internal magnetic fields. For example, by estimating the temporal variations of the $p$-mode frequency varying with the solar cycle, \cite{2010ApJ...718L..19F} observed a quasi-biennial variation in the natural oscillation frequencies in the solar interior.  Based on the their analysis results, they pointed out that the periodicity around 2 years is distinct and separate from the classical 11-year Schwabe cycle. An alternative explanation with the potential to reproduce some of the temporal features (for instance, the intermittency and the variable periodicity) of solar QBOs is the instability of the $m=1$ magnetic Rossby waves in the solar tachocline \citep{2010ApJ...724L..95Z}. They found that, at the base of the convection zone, if the magnetic field strength is strong with an order of $10^5$ G, the instabilities will produce the periodicities in the range of two years. Moreover, \cite{2010ApJ...724L..95Z} and \cite{2017ApJ...845..137G} found that the magnetic Rossby waves could be also applied to examine the north-south asymmetry of solar magnetic features and their periodicities. 

In the past, solar magnetic activities were found to exhibit a significant hemispheric asymmetry in the temporal profile \citep{1971SoPh...20..332W, 1996SoPh..166..201A, 2019SoPh..294..142C}. Using various magnetic indices and proxies such as the numbers or areas of sunspot, filaments and prominences, plage areas, solar flares, coronal bright points, and coronal mass ejections, the north-south asymmetry of solar activity indices in different layers of the solar atmosphere has been widely studied and examined \citep{2005A&A...431L...5B,2013ApJ...768..188C, 2013ApJ...779....4B, 2014SSRv..186..251N, 2016Ap&SS.361..208J, 2018MNRAS.473.1596B}. By studying the observational data of the brightness of the coronal green line (530.3 nm) and sunspot activity indices over the period of 1939-2001, \cite{2008SoPh..247..379B} found that solar QBOs in the asymmetry time series of activity indicators are more obviously than the QBOs existed in the activity indicators themselves. Moreover, the relative QBO power shown in the north-south asymmetry of the considered indicators exhibits an anti-correlation with the north-south asymmetry of the given indicators themselves. Later, the spatial and temporal distribution of the hemispheric asymmetry of sunspot groups in the time interval 1874-2009 was studied by \cite{2011NewA...16..357B}. The main conclusion is that an increase in the hemispheric asymmetry of the given data set is accompanied by a decrease in the amplitude of solar QBOs. They thus suggested that,  to a great extent, solar magnetic activity is generated independently (i.e., weak coupling) in northern and southern hemispheres. Moreover, it is still controlled by the rules of the rotational behavior and the meridional flow in each of the two hemispheres. \cite{2018SoPh..293..124M} applied the multichannel singular spectrum analysis method to study the intermediate periodicities of the coronal green emission line (530.3 nm) for the time period from 1944 to 2008. They found that the north-south asymmetric distribution, i.e., the uneven latitudinal distribution of solar QBOs, could be considered as a fundamental, but puzzling, characteristic of solar magnetic activity. Using the simulations from the flux transport dynamo model, \cite{2019A&A...625A.117I} analyzed the sunspot areas to investigate the space-time behavior of the QBO signal at low latitudes. They found that the signal of solar QBOs is not only present in the full disk, but also shown in the sunspot areas in the two hemispheres. Furthermore, the QBOs show different spatial and temporal behaviors, suggesting slightly decoupled solar hemispheres. The QBO signal in the low atmosphere of the Sun is intermittent and in-correlation (i.e., in-phase) with the sunspot activity, surfacing when the solar magnetic activity arrives at its maximum value.

Based on the data sets of the numbers or areas of sunspot, the H$\alpha$ flare index, and the coronal emission line intensity, the north-south asymmetry of solar QBOs were widely studied by previous works. However, these phenomena, which are related to the sunspots or flares, are usually considered as the magnetic activity at low latitudes ($\leq50^\circ$). That is, the statistical and observational characteristics of the north–south asymmetry of solar QBOs at high latitudes ($\geq60^\circ$) are rarely investigated. According to \cite{2002PASJ...54..787L,2004PASJ...56L..49L},  polar faculae are usually considered as a typical indicator of solar magnetic activity at high latitudes, and they are found to be anti-correlated with the magnetic activity of the sunspot-related indices. In the present work, the monthly counts of the polar faculae in the northern hemisphere and the southern hemispheres, recorded on the sunspot sketches designed by the National Astronomical Observatory of Japan, are applied to study the hemispheric distribution of solar QBOs at high latitudes. In the following, a brief introduction of the polar faculae data and the data analysis technique is shown in Section 2, and the detailed statistical results based on the time-frequency analysis methods are presented in Section 3. Section 4 gives the main conclusions and discussions of this work.

\section{Observational Data}

\subsection{Polar faculae}
\label{sec:data} 

\begin{figure}
   \includegraphics[width=\columnwidth]{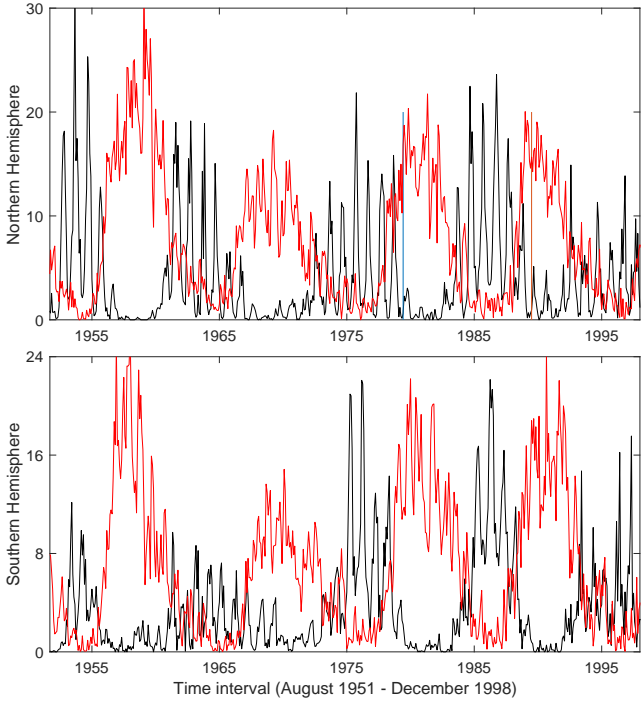}
    \caption{The monthly time series of the polar faculae (indicated by the black lines) in the northern hemisphere (upper panel) and the southern hemisphere (lower panel) , respectively, during the period from August 1951 to December 1998. The monthly time-series of the sunspot numbers are also shown in red.  The amplitude of the hemispheric sunspot numbers are divided by 5 for visual comparison.}
    \label{fig:figure1}
\end{figure}

Faculae are small-scale bright structures that are usually appeared and observed in the solar limb (mostly greater than $60^\circ$). Naturally, they have a close connection with the magnetic field concentrations. To improve the knowledge on the faculae at polar regions, \cite{2016MNRAS.460..956Q} studied the magnetic properties of the spatially polar faculae. They found that the polar faculae are formed by hot plasma with low line-of-sight velocity and that the hot wall effect might allow the detecting photons, which comes from deeper layer, appeared closer to the limb of the Sun.

The long-term data sets applied in this work were taken from the Mitaka of the NAOJ\footnote{http://solarwww.mtk.nao.ac.jp/en/db\_faculae.html}, which provided the monthly time series of the polar faculae in the northern and southern hemispheres. The time interval is from August 1951 to December 1998, covering the solar cycles 19-22. For details, please refer to the papers those summarized the long-term observing investigations of the Sun at the NOAJ \citep{1993RNAOJ...2..403I, 1998ASPC..140..483S, 2013JPhCS.440a2041H}.

Figure 1 displays the monthly time series of the polar faculae (indicated by the black lines) in the northern hemisphere (upper panel) and the southern hemisphere (lower panel) , respectively, during the period from August 1951 to December 1998. From this figure, one can easily see that the time series of the polar faculae do not asynchronously peak in the two hemispheres, suggesting magnetic activity at high latitudes should apparently be phase asynchronous and amplitude asymmetric on both hemispheres.

To better show the phase relationship between the polar faculae (representing the high-latitude solar activity) and the sunspot numbers (representing the low-latitude solar activity), the monthly time series of the sunspot numbers (indicated by the red lines) in the northern hemisphere (upper panel) and the southern hemisphere (lower panel) are also shown. Here, the monthly time series of the sunspot numbers in the two hemispheres were freely taken from the website of the WDC-SILSO (World Data Center, Sunspot Index and Long-term Solar Observations), Royal Observatory of Belgium, Brussels\footnote{http://www.sidc.be/silso/datafiles}. Similar to the polar faculae, the time interval of the sunspot numbers used here is also from August 1951 to December 1998. It should be noticed that the amplitude of the hemispheric sunspot numbers is divided by 5 for better visual splendour.

As the presence of the polar faculae is immediately related to the inner magnetism of the Sun, the polar faculae follow an 11-year cycle, as shown in Figure 1. However, the progress of the apparent cycle is anti-correlated to the normal sunspot cycle. That is to say, when the level of the sunspot activity increases, the level of the facular activity decreases, i.e., the faculae regions begion to migrate towards the two poles until they vanish when the biggest value of the sunspot activity is reached \citep{1996SoPh..163..267M, 2003SoPh..214...41M}. Here, we only show the different temporal behaviors between the polar faculae and the sunspot numbers, their detailed quantitative relationship is beyond the scope of this work.

\subsection{Synchrosqueezed wavelet transform}
\label{sec:data} 

The synchrosqueezed wavelet transform (SWT), which was first introduced by \cite{daubechies2011synchrosqueezed}, is a relatively novel time-frequency analysis technique. This method is also called as an empirical mode decomposition (EMD)-like tool. At present, this technique has been applied to the fields of solar physics \citep{2017ApJ...845...11F}, geophysics \citep{2018GeopP..66.1358X}, civil engineering \citep{rafiei2018novel}, and achieved good results.

On the basis of the continuous wavelet transform (CWT), the time-frequency resolution and  the anti-noise ability of the SWT are obviously improved. For a given signal $s(t)$, the wavelet coefficients $W_s(p,q)$ calculated by the CWT could be described as:

\begin{equation}
W_{s}(p,q)=\int_{-\infty}^{+\infty} s(t) p^{-1 / 2} \psi^*\left(\frac{t-q}{p}\right) d t
\end{equation}

here, $p$ and $q$ are the scale factor and the shift factor. $\psi$ denotes the mother wavelet, and $\psi^*$ is its complex conjugate.

The calculation principle of the SWT technique is to concentrate the energy in the time-frequency map into the instantaneous frequency. The corresponding frequency $\omega_s(p,q)$ of each scale can be obtained by the derivative of the wavelet coefficients with respect to the time,

\begin{equation}
\omega_{s}(p,q)=\frac{-i}{W_{s}(p,q)} \frac{\partial}{\partial q} W_{s}(p,q)
\end{equation}

\begin{figure}
    \includegraphics[width=\columnwidth]{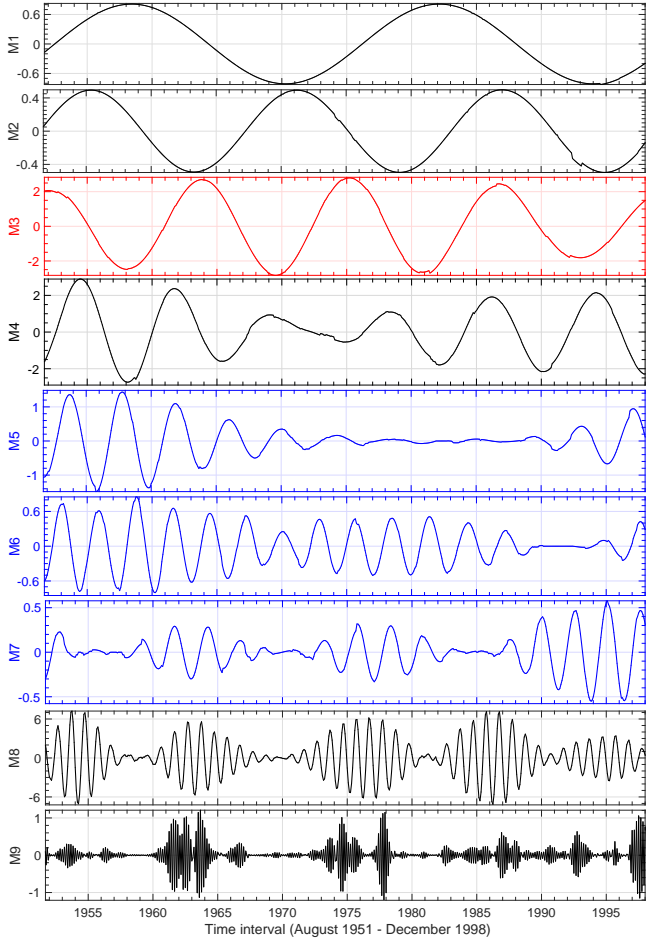}
    \caption{The decomposed $M_i$ modes ($i$ values from 1 to 9) of the polar faculae in the northern hemisphere obtained by performing the SWT technique. Panel 3 (filled with red) is the 11-year mode, and panels 5-7 (filled with blue) are the QBO components.}
    \label{fig:figure2}
\end{figure}

\begin{figure}
    \includegraphics[width=\columnwidth]{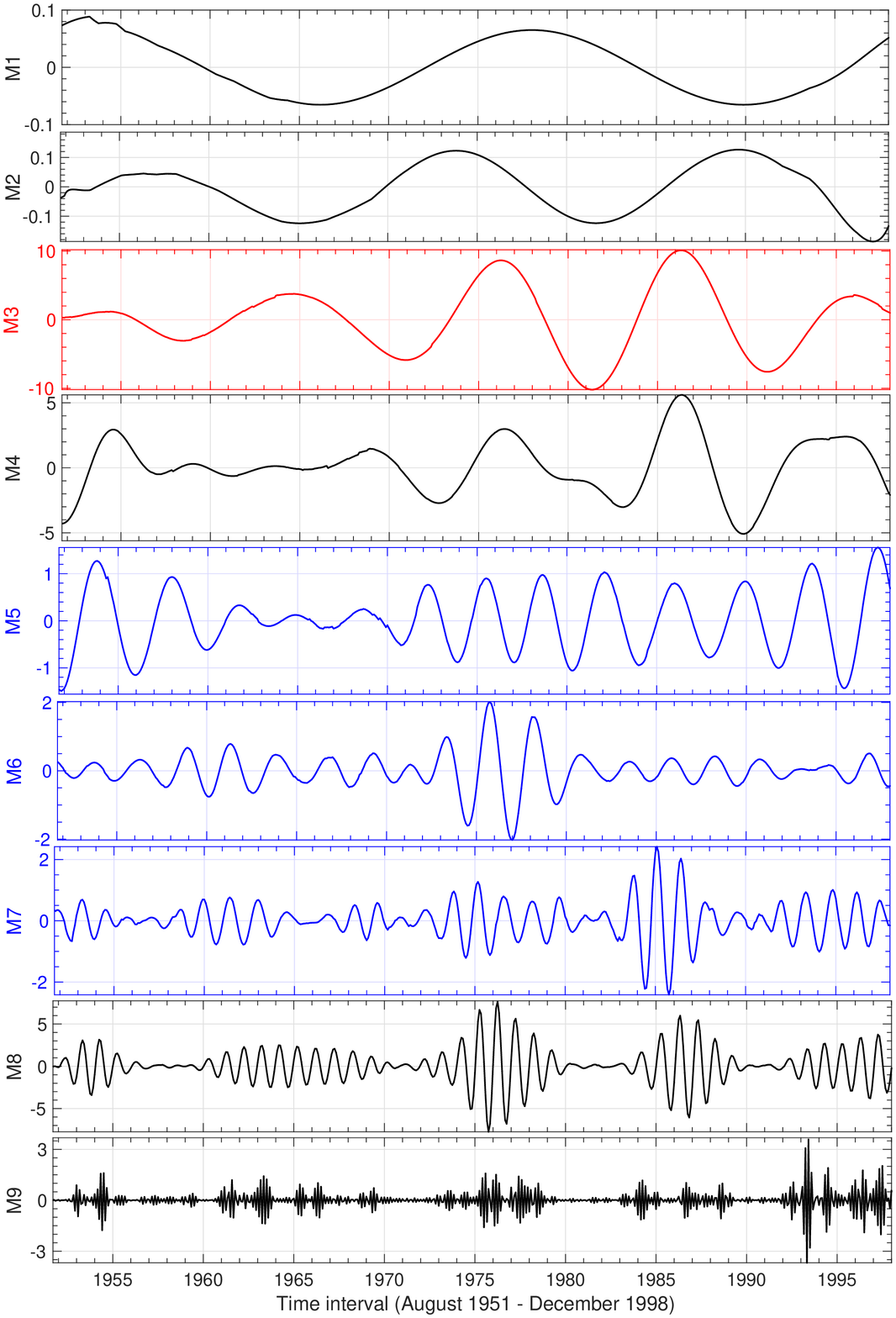}
    \caption{The decomposed $M_i$ modes ($i$ values from 1 to 9) of the polar faculae in the southern hemisphere obtained by performing the SWT technique. Panel 3 (filled with red) is the 11-year mode, and panels 5-7 (filled with blue) are the QBO components.}
    \label{fig:example_figure}
\end{figure}

Actually, the above parameters $p$, $q$, and $\omega$ are discretized, namely, $W_x(p,q)$ is calculated only at a discrete $p_k$ with $p_k-p_{k-1}=(\delta p)_k$. Then, the synchrosqueezed transform  $T_s(\omega,q)$ is eastimated only at the center frequency $\omega_n$ of successive frequency bins $[\omega_n-\delta \omega/2,\omega_n+\delta \omega/2]$ with $\delta \omega=\omega_n-\omega_{n-1}$, by summing up different contributions:

\begin{equation}
T_{s}\left(\omega_{n}, q\right)=\frac{1}{\delta \omega} \sum_{p_{k}} W_{s}\left(p_{k}, q\right) p_{k}^{-3 / 2}(\delta p)_{k}
\end{equation}

The fully discretized estimate of the SWT can be calculated by $\tilde{T}_{\tilde{s}}\left(\omega_{n}, t_{m}\right)$, where $t_m$ denotes the discrete sampling localization in the time domain, and $m$ denotes the total numbers of the samples in the considered time series $s(t)$. 

Each mode $M_i$ of the time series $s(t)$ is reconstructed through an inverse synchrosqueezed wavelet transform over a small frequency band $ l \in L_{k}\left(t_{m}\right)$,

\begin{equation}
M_{i}\left(t_{m}\right)=2 C_{\varphi}^{-1} \Re\left(\sum_{\ell \in L_{k}\left(t_{m}\right)} \tilde{T}_{\tilde{s}}\left(w_{\ell}, t_{m}\right)\right)
\end{equation}

where $C_{\varphi}$ denotes a constant which comes from the selected wavelet, $\Re$ is the real part of the wavelet time-frequency representation, and $\tilde{T}_{\tilde{s}}\left(\omega_{n}, t_{m}\right)$ is the discretized version of $T_s(\omega_n,b)$. A detailed description (such as the calculation principle, robustness properties, and so on) of the SWT algorithm can be found in papers of   \cite{daubechies2011synchrosqueezed}, \cite{thakur2013synchrosqueezing}, and \cite{2019SGeo...40.1185X}.

Naturally, both the EMD and the SWT can be applied to decompose a given time series and can extract the intrinsic mode functions. However, unlike to the EMD technique, the SWT algorithm has a firm theoretical foundation, so the intrinsic modes extracted by the SWT technique can be explained from a physical point of view. Moreover, the EMD technique is sensitive to the noise, while the adaptive and the invertible transform processes make the SWT method more robust to noise. As attested by \cite{2017ApJ...845...11F}, the SWT technique can locate the frequency components with a high spectrum resolution, and produces a sharper time-frequency map of the time series. On the basis of these two advantages, we choose the SWT to decompose the time series of the polar faculae in the two hemispheres.

\section{Analysis of Results}

\subsection{Period modes of hemispheric polar faculae}

By applying the SWT technique, nine period modes are decomposed for the polar faculae in each of the two hemispheres. The decomposed results are shown in Figures 2 (northern hemisphere) and 3 (southern hemisphere). Here, the bump wavelet is chosen, and the bin value of $\delta \omega$ is set to 4.

Based on the SWT technique, the values of the mean periodicity (in unit of years) of each mode for the polar faculae in the two hemispheres are calculated. The results for the northern and southern hemispheres are listed in the second and third columns of Table 1. The corresponding error ranges of each intrinsic mode are also evaluated through the bin values. From Table 1, it is found that each time series presents one period mode ($M_1$) representing the nearly 22-year magnetic cycle; one period mode ($M_3$) representing the nearly 11-year Schwabe cycle; and one period mode ($M_9$) representing the sampling interval (one month) of the data sets.

The periodic value of around 16 years ($M_2$) appears, respectively, in the northern hemisphere (16.01 years) and the southern hemisphere (15.87 years), and the length of this value is about 1.5 times (10.7 $\times$ 1.5= 16.05) as long as the 11-year Schwabe solar cycle. Moreover, this periodicity can be taken as the 17-year solar extended activity cycle, covering from the maximum time of a certain solar cycle to the minimum time of the succedent solar cycle. For example, by examining the latitude distribution variation of the coronal magnetic activity during the period 1944-1974, \cite{1983A&A...120L...1L} found that the the whole evolution process of the magnetic activity occurring in the corona spreads over a time interval of around 17 years. Furthermore, they provided a strong evidence that the duration of the magnetic activity in the solar atmosphere is longer than the time interval (i.e., 11 years) between two consecutive solar cycles. \cite{1980ApJ...239L..33H} reported that the length of the solar extended activity cycle is greater than the classical 11-year Schwabe solar cycle, and the torsional oscillations provided a solid evidence for an even longer extended cycle. \cite{2010ApJ...716..693R} found that the latitude varying with the time plots (the so-called butterfly diagrams) of the coronal emission variation exhibit a zone of enhanced brightness that occurs near the two poles after the maximum time of a solar cycle and then migrates toward to the lower latitudes, and a bifurcation occurs at the solar minimum with one branch continuing to migrate equator-ward with the sunspots of the new solar cycle and the other branch heading back to the solar poles. They thought that their resulting patterns liken to those seen in the temporal variation of the torsional oscillations and taken as an solid evidence for the existence of the 17-year extended solar cycle.

The periodicity of around 6-7 years ($M_4$) might be connected with the 22-year Hale cycle or the magnetic field reversal, because it is the third harmonic of the 22 years. By investigating the temporal variation of coronal global rotation in the 10.7 cm solar radio flux (2800 MHz), \cite{2017ApJ...841...42X} obtained a typical periodicity of 6.6 years in the periodic length of the coronal rotation, and this periodicity was interpreted as the third harmonic of the Hale cycle (22 years). This periodicity was also found and reported by \cite{1995SoPh..158..173J} and \cite{2016SoPh..291.3485J} whose studies focused on the equatorial rotation rate of solar magnetic structures (sunspot groups). However, the physical origin of the periodicity around 6-7 years is still unknown, further works are needed to deal with this question.

For the period mode ($M_8$) of both data sets, the periodic value is about one year, which is mostly related to annual-variation signal. Although the periodicity around one year has been reported in many solar indices such as the solar constant (total solar irradiance), the solar mean magnetic field, and the solar full-disk magnetic field \citep{2016AJ....151...76X}, but its exactly physical mechanism is still doubtful, as pointed by \cite{2009SoPh..257...61J}. That is to say, it is hard to rule out the possibility that it is not caused by the influence of the seasonal effects of the Earth.

From the above analysis, the period mods of polar faculae in the two hemispheres are found to mainly consist of the 22-year magnetic cycle ($M_1$ and $M_4$), the 17-year extended activity cycle ($M_2$), the 11-year Schwabe cycle ($M_3$), the annual-variation signal ($M_8$), and the sampling interval of the data sets ($M_9$). The rest of the period modes (from $M_5$ to $M_7$) can be considered as solar QBOs existing in the high-latitude solar activity.

\begin{table}
	\centering  
	\caption{The mean periodicities (in unit of years)  and their error ranges of the nine period modes which are decomposed from the polar faculae in the northern hemisphere and the southern hemisphere.}.
	\begin{tabular}{ccccc}
		\hline\hline
	 Mode    & Northern hemisphere      & Southern hemisphere\\
		\hline
		M1	  & $23.86^{+1.081}_{-1.034}$ 		& $24.01^{+1.464}_{-1.380}$	                      \\
		M2   & $16.01^{+0.726}_{-0.694}$ 		& $15.87^{+0.967}_{-0.912}$			     \\
		M3	  & $10.81^{+0.532}_{-0.469}$ 	         & $10.70^{+0.652}_{-0.615}$	             \\
		M4	  & $7.095^{+0.322}_{-0.308}$           & $6.481^{+0.395}_{-0.372}$		     \\
		M5	  & $4.052^{+0.184}_{-0.176}$ 		 & $3.724^{+0.227}_{-0.214}$                      \\
		M6	  & $2.843^{+0.129}_{-0.123}$ 	          & $2.535^{+0.155}_{-0.146}$		     \\
		M7   & $2.544^{+0.115}_{-0.110}$             & $1.445^{+0.088}_{-0.083}$	             \\
		M8	  & $1.051^{+0.048}_{-0.047}$            & $0.984^{+0.060}_{-0.057}$		    \\
	        M9	  & $0.170^{+0.005}_{-0.004}$            & $0.167^{+0.010}_{-0.009}$		    \\
		\hline
	\end{tabular}
	\label{T_ccx_subregion}
\end{table}

\subsection{Hemispheric distribution of solar QBOs}

\begin{figure}
    \includegraphics[width=\columnwidth]{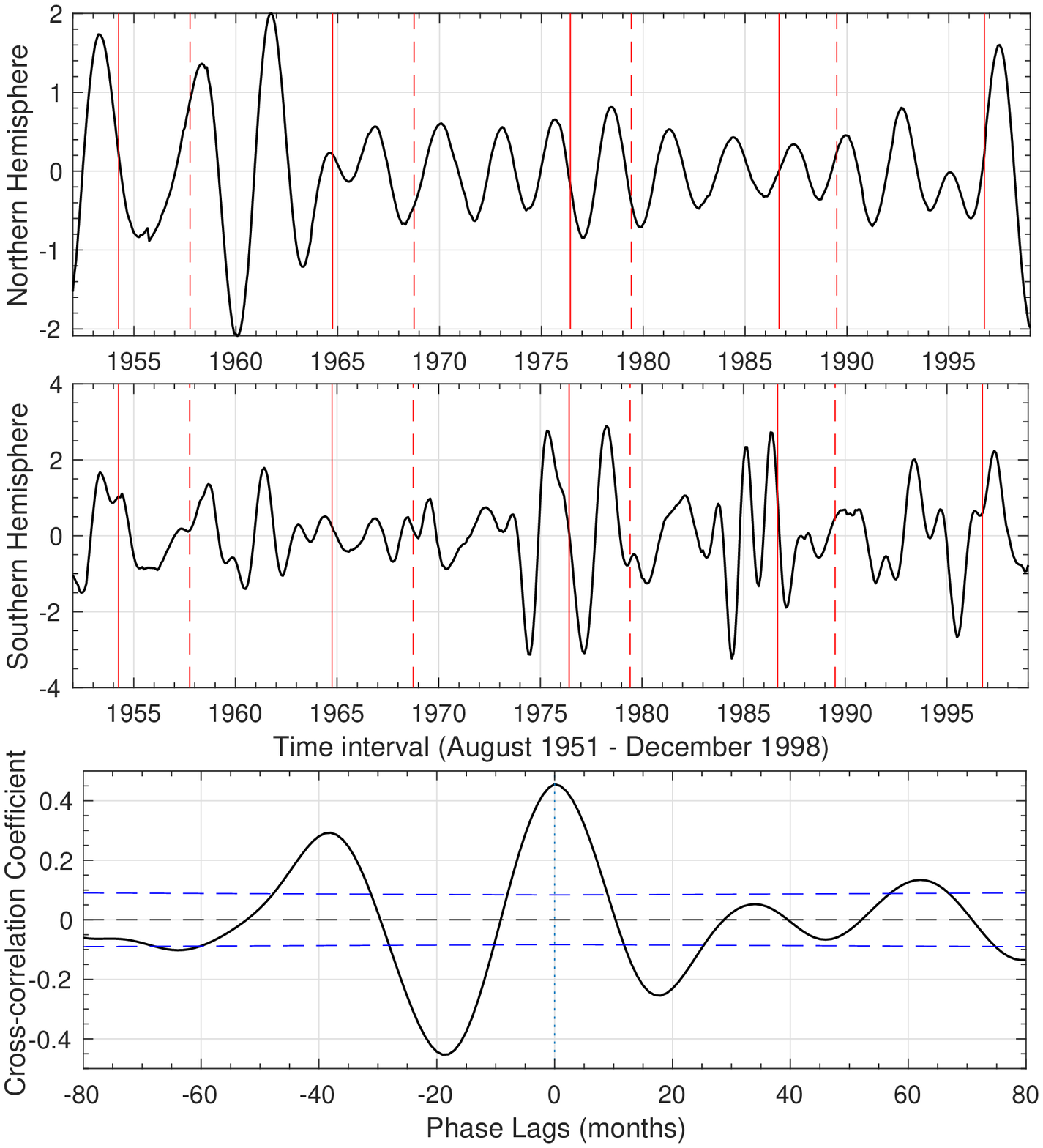}
    \caption{Top panel: solar QBOs ($M_5$ to $M_7$) of the polar faculae in the northern hemisphere. Middle panel: solar QBOs ($M_5$ to $M_7$) of the polar faculae in the southern hemisphere. Here, the minimum and maximum times of normal solar cycles are indicated by the vertical solid and dashed red lines in the top and middle panels, respectively. Bottom panel: the cross-correlograms between the  northern QBOs and the southern QBOs, and the 95\% confidence levels are indicated by the blue dashed lines.}
    \label{fig:example_figure}
\end{figure}

\begin{figure}
    \includegraphics[width=\columnwidth]{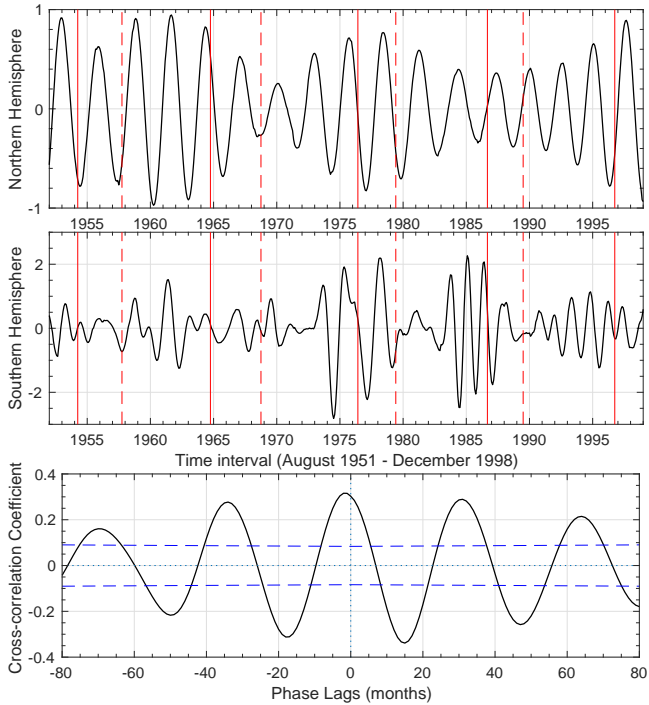}
    \caption{Top panel: solar QBOs ($M_6$ to $M_7$) of the polar faculae in the northern hemisphere. Middle panel: solar QBOs ($M_6$ to $M_7$) of the polar faculae in the southern hemisphere. Here, the minimum and maximum times of normal solar cycles are indicated by the vertical solid and dashed red lines in the top and middle panels, respectively. Bottom panel: the cross-correlograms between the  northern QBOs and the southern QBOs, and the 95\% confidence levels are indicated by the blue dashed lines.}
    \label{fig:example_figure}
\end{figure}

Here, the full contribution of solar QBOs of high-latitude solar activity in each hemisphere is reconstructed by summing up the intrinsic modes having typical periodicities in the range from 1.3 years to 4.0 years. Although the SWT technique is a relatively novel technique that can be applied to clearly separate each mode of a given time series, it also requires the signal at a certain timescale have a higher spectrum intensity than its surrounding timescales. That is, only the mode extracted from the original time series is an intrinsic mode, it can be taken as a candidate component of the QBOs at high latitudes. That is way it is needed to compute the full contributions of the QBOs in each hemisphere by summing up the modes with timescales in the interval from 1.3 years to 4.0 years.

 It should be pointed out that solar QBOs are usually defined as the oscillations with periodicity between 1.5 years and 3.5 years. However, the average periods of the mode $M_5$ are 4.052 years and 3.724 years are, respectively, found in the northern hemisphere and the southern hemisphere. Therefore, to reconcile the intention of isolating QBOs, the extension of the upper limit of the QBO components to 4.0 years is applied. 

Figure 4 shows the time profile of solar QBOs at high latitudes, the top and the middle panels correspond the northern and the southern hemispheres, respectively. Both of them are obtained through the sum of three modes ($M_5$ to $M_7$) whose typical periods locate the range of 1.3-4.0 years. From these two panels, the reconstructed QBOs are coherent in the two hemispheres, although the mean amplitude is not identical for the two time series. To show the temporal evolution of solar QBOs varying with the solar cycle, in the top and middle panels, the minimum and maximum times of the normal solar cycles are indicated as the vertical solid and dashed red lines, respectively. During the sunspot maxima, solar QBOs at high latitudes seemingly have a low amplitude, for example in cycles 20 and 21. In the sunspot minima, they seemingly have a high amplitude, for instance during cycles 19 and 23. However, this feature is not clear and not typical for the all solar cycles studied here. Moreover, this figure clearly shows that the temporal (phase) and the spatial (amplitude) variations of solar QBOs occur unevenly on both hemispheres. At most of the time points, the amplitude of solar QBOs in the southern hemisphere is obviously greater than that in the northern hemisphere.

To understand the hemispheric difference of the QBOs of solar magnetic activity at high latitudes, we calculate the ratio between the mean-square amplitude of solar QBOs and that of the 11-year Schwabe cycle mode ($M_3$) in the two hemispheres, respectively. The values are found to be 14.38\% for the northern polar faculae and 4.84\% for the southern polar faculae, implying that the importance of solar QBOs at high latitudes with respect to the 11-year Schwabe cycle mode in the two hemispheres is not identical.

According to the top and middle panels of Figure 4, it can be seen that the temporal and the spatial evolutions of the hemispheric QBOs in the polar faculae obviously behaves differently, implying that the QBOs of the polar faculae should be (phase) asynchronous and (amplitude) asymmetric in the two hemispheres. Here, the cross-correlation analysis method, which is usually to examine the intercorrelation between the two time series, is applied to study the phase relation of hemispheric QBOs of the polar faculae. The resulting cross-correlograms are shown in the bottom panel of Figure 4. The abscissa denotes the phase lags (in the range between -80 and 80 months) of the northern hemisphere with respect to the southern hemisphere. The negative (positive) values representing the backward (forward) shifts (namely, the northern hemisphere peaks later (earlier) in time). The 95\% confidence levels of the cross-correlation coefficients, which is used to estimate the statistical significant, are shown as the blue dashed lines.

When the two time series has no phase shift (i.e., the phase lag is zero), the cross-correlation coefficient is 0.46 and it is the largest one. That is, from a global point of view, the hemispheric QBOs at high latitudes are positive correlation. As displayed in the bottom panel, with the range of the shifts between -80 and 80 months, there are four local maxima in the cross-correlograms. The values of the cross-correlation coefficient peak at local maxima of 0.29, 0.46, 0.05, and 0.13, when the phase lags between the two time series are -38, 0, 34, and 62 months, respectively. The mean length between each two neighboring local maxima is 33.3$\pm$5.0 months (2.78$\pm$0.42 years).  When their phase shifts are -64, -19, 18, and 46 months, the cross-correlation coefficients arrive at local minima with the values of -0.10, -0.45, -0.25, and -0.07, respectively. The mean length between each two neighboring local minima is 36.7$\pm$8.5 months (3.06$\pm$0.71 years). That is to say, the phase relationship of solar QBOs of the polar faculae in the two hemispheres is not a simply way, showing no systematic regularity.

As mentioned above, solar QBOs are usually defined as the oscillations with the typical periodicities between 1.5 years and 3.5 years. So it is also needed to examine the hemispheric distribution of solar QBOs when the $M_5$ is not taken into account. Now, by summing up the modes $M_6$ and $M_7$ in each solar hemisphere, the hemispheric QBOs of the polar faculae are reconstructed. The resulting QBOs in the northern and southern hemispheres are, respectively, displayed in the top and middle panels of Figure 5. As shown in the two panels, the temporal (phase) and the spatial (amplitude) variations of solar QBOs at high latitudes occur unevenly on both hemispheres.

The cross-correlation analysis method is also applied to examine the phase relation of the hemispheric QBOs (the $M_5$ is not taken into account) shown in Figure 5. The resulting cross-correlograms are shown in the bottom panel of Figure 5. By comparing the two cross-correlograms, one can found that the profile shown in Figure 5 is more smooth and prefect than that shown in Figure 4.  When the the phase lag is zero, the cross-correlation coefficient is 0.30, but it is not the largest one. When the phase lag is -2 months, the cross-correlation coefficient reaches its maximum with a value of 0.32. That is, from a global point of view, the hemispheric QBOs at high latitudes are positive correlation, but the southern hemisphere peaks 2 months earlier than the northern hemisphere. Within the whole range of the phase shifts between -80 and 80 months, there are five local maxima (0.16, 0.28, 0.32, 0.29, and 0.22) in the cross-correlograms, and the  corresponding phase lags are -70, -34, -2, 31, and 64 months. The mean length between each two neighboring local maxima is 33.5$\pm$1.7 months (2.79$\pm$0.14 years).  When their phase shifts are -50, -18, 15, and 47 months, the cross-correlation coefficients arrive at local minima with the values of -0.22, -0.31, -0.34, and -0.26, respectively. The mean length between each two neighboring local minima is 33.5$\pm$0.58 months (2.79$\pm$0.05 years). That is, when the mode $M_5$ in each hemisphere is not taken into account, the phase relationship of the hemispheric QBOs of the polar faculae becomes simple.

\begin{figure}
    \includegraphics[width=\columnwidth]{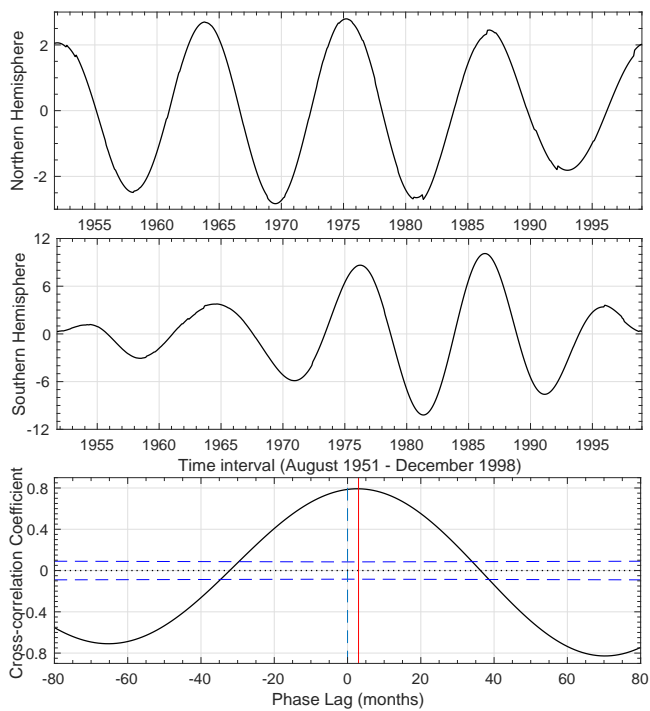}
    \caption{Top panel: the 11-year mode ($M_3$) of the polar faculae in the northern hemisphere. Middle panel: the 11-year mode ($M_3$) of the polar faculae in the southern hemisphere. Bottom panel: the cross-correlograms  of the 11-year mode between the two hemispheres, and the 95\% confidence levels are indicated by the blue dashed lines.}
\end{figure}

\subsection{Comparison with the 11-year mode}

To compare the hemispheric distribution of the polar faculae at different timescales, we also investigate the phase relation of the 11-year mode of the hemispheric polar faculae through the cross-correlogram analysis.  The temporal variations of the 11-year mode of the polar faculae in the northern hemisphere (top panel) and the southern hemisphere (middle panel) are displayed in Figure 6. It is clearly shown that the temporal (phase) and the spatial (amplitude) evolutions of the 11-year mode in the two hemispheres behave differently in the studied time range, suggesting a slight decoupling of the high-latitude solar activity between the two hemispheres.

The cross-correlation analysis is also performed to examine the phase relation of the hemispheric polar faculae at the 11-year timescale, and the result is displayed in the bottom panel of Figure 6. Similar to the QBOs studied in Figures 4 and 5, the phase shifts of the 11-year mode are also chosen between -80 and 80 months.

As this panel displays, the phase shift of the 11-year mode between the two hemispheres is estimated to be three months when the cross-correlation coefficient reaches its maximum with a value of 0.79. The biggest correlation coefficient (0.79) is obviously far from the 95\% confidence lines (the blue dashed lines shown in this figure), indicating that the obtained best correlation is highly statistical significant, and such a phase leading can bring out an increase in the asynchronous behavior of the hemispheric distribution of solar magnetic activity at high latitudes.

As shown in Figures 4, 5, and 6, within the cross-correlation analysis results displayed here, it is easily found that solar QBOs of high-latitude magnetic activity exhibit a complex phase relation in the two hemispheres, i.e., they do not have any regular pattern. For the time interval considered here, the 11-year mode (mode 3 shown in Figure 2) of high-latitude magnetic activity in the northern hemisphere peaks three months earlier than the 11-year mode (mode 3 shown in Figure 3) of high-latitude magnetic activity in the southern hemisphere. In other words, the spatial distribution of solar QBOs in the two hemispheres is different from that of the Schwabe cycle mode (i.e., the 11-year mode) in the two hemispheres.

\section{Conclusions and Discussions}

By utilizing of the monthly values of the polar faculae recorded from August 1951 and December 1998, the spatial distribution of the quasi-biennial oscillations in the high-latitude solar magnetic activity was detailed studied.  Firstly, the SWT technique was applied to extract the periodic modes of the polar faculae and to calculate their typical periodicities. Then, the hemispheric phase relationship of solar QBOs (one is from $M_5$ to $M_7$, and the other one is $M_6$ and $M_7$) was examined by the cross-correlogram analysis. At last, to compare the hemispheric distribution of the polar faculae at different timescales, the phase relationship of the 11-year mode (the Schwabe solar cycle) of the polar faculae in the two hemispheres was also studied.

According to the SWT analysis technique, the monthly polar faculae in each hemisphere were decomposed into nine modes.  The typical period modes of the polar faculae in the two hemispheres are found to mainly consist of the 22-year Hale cycle, the 17-year extended activity cycle, the 11-year Schwabe cycle, the solar QBOs, and the annual-variation signal. However, the values of nine period modes on both hemispheres are not exactly the same, implying that the solar poloidal fields at high latitudes operating in the two hemispheres exhibit different behaviors.

 As is known to all, the spatial and temporal distributions of solar magnetic features at different layers of the Sun, such as sunspots or sunspot groups, flares, filaments, surface magnetic fields, as well as the brightness of the corona green line, during a certain solar cycle are connected with the meridional circulation, which could be taken as a component of the sub-photospheric flows. For instance, \cite{2010ApJ...716..693R} found that the butterfly diagrams of the coronal green emission exhibit a zone of enhanced brightness that occurs at the two poles after the maximum time of a solar cycle and migrates toward to low-latitude belts, and a bifurcation occurs at the minimum time of a solar cycle with one branch continuing to migrate equator-ward with the sunspots of the new solar cycle and the other branch heading back to the two poles. The time range of the whole dynamic process is about 17-18 years. Therefore, the 17-year periodicity of the polar faculae found in this work, providing a scenario that the dynamical behavior of solar activity at high latitudes might be related to the sub-photospheric flows and the global meridional circulation.

For the reconstructed QBOs of the hemispheric polar faculae, they are coherent in the two hemispheres, but the mean amplitude is not identical for the two time series. During the sunspot maxima, solar QBOs at high latitudes seemingly have a low amplitude (cycles 20 and 21). In the sunspot minima, they seemingly have a high amplitude (cycles 19 and 23). However, this feature is not clear and not typical for the all solar cycles studied in this work. For the low-latitude solar activity indicators, such as the numbers the areas of sunspot, the time profile of solar QBOs have a high amplitude in the minimum time of each solar cycle, while they vanish in the maximum time of the solar cycle. That is to say, the temporal variations of solar QBOs at high latitudes are different from those at low latitudes. Such difference can be interpreted as the phase asynchrony between the polar faculae and the sunspot numbers. As pointed out by \cite{2006SoPh..236..185L}, the polar faculae have an anti-correlation with the sunspot-related indices with a phase lag of around 50-71 months. As the activity level of sunspots and sunspot groups increases, the faculae regions begin to slowly migrate to the two poles of the Sun until they vanish when the maximum value of the sunspot activity is reached. 

The temporal variation of solar QBOs at high latitudes occurs unevenly in the two hemispheres, because the importance of solar QBOs at high latitudes with respect to the Schwabe cycle mode in the two hemispheres is not identical, one is 14.38\% (northern hemisphere) and the other is 4.84\% (southern hemisphere). Using the cross-correlation analysis method, we found that the phase relationship of solar QBOs (by summing up modes from $M_5$ to $M_7$) of the hemispheric polar faculae is very complex, showing no systematic regularity. When the $M_5$ in each hemisphere is not taken into account, the phase relationship of the hemispheric QBOs of the polar faculae becomes simple, because the profile of the cross-correlograms is more smooth and prefect. From a global point of view, the hemispheric QBOs at high latitudes are positive correlation, but the southern hemisphere peaks two months earlier than the northern hemisphere. However, for the 11-year period mode, the northern hemisphere peaks three months earlier than that in the southern one. That is to say, the spatial distribution of the hemispheric QBOs (no matter the mode $M_5$ is taken into account or not) is different from that of the 11-year mode in the two hemispheres. Actually, the magnetic field strength and the parameters of the differential rotational on both solar hemispheres vary with the time depending on the phase and the amplitude during a given solar cycle, bringing out different growth rates of the magnetic field in the two hemispheres.

Now, we discuss the statistical results from a viewpoint of the dynamical variation of the Sun. According to the solar dynamo model proposed by \cite{1961ApJ...133..572B} and \cite{1964ApJ...140.1547L}, at relatively low latitudes ($\leq50^\circ$), magnetic flux emergence is generated in the form of solar active regions and sunspot groups, and the temporal variation of the large-scale magnetic field is considered to be caused by the surface transport and dispersal of the magnetic flux inside the active regions \citep{1996SoPh..163..267M, 2003SoPh..214...41M}. In fact, the movement of solar magnetic field from low-latitude belts to high-latitude belts forms the poloidal fields in the two poles \citep{1981SoPh...74..131H}. From the high-spatial observations, solar magnetic fields at high latitudes ($\geq60^\circ$) might come from two parts: the first one is from the flux emergence of the magnetic activity from the interior of the Sun and the second one is from the polewards drift of the magnetic field at low latitudes. Therefore, the temporal and spatial distribution, such as the periodic values of solar magnetic activity and the hemispheric asymmetric distribution at high latitudes is thought to have little or no relation with that at low latitudes.
 
Based on the solar dynamo theories, there are several possibly theoretical models applied to explain the hemispheric asymmetric distribution of solar magnetic activity cycle. For example, by using the mean field theory, \cite{2017ApJ...835...84S}  investigated the relationship between the dipole and the quadrupole modes of solar magnetic field. They found that, in the solar magnetic activity cycle, two different attractors might exist. Moreover, their model can be also used to explain the amplitude and phase asymmetry of solar magnetic activity and the polar magnetic field reversal. A recent work proposed by \cite{2018A&A...618A..89S} showed that the absolute asymmetric variation of solar magnetic activity can be interpreted as the superposition of an excited dipolar-type component and a linearly damped, but randomly excited quadrupolar-type component. Their model was made on the basis of an updated and developed Babcock-Leighton dynamo model. The main results derived from our statistical analyses may help to better reveal and understand the physical origin of the spatial distribution of solar QBOs at high latitudes and prediction of their dynamical behavior. In addition, our analysis of solar QBOs at high latitudes might be directly connected with other astrophysical fields, for instance, the QBOs of solar neutrino fluxes, the modulation of cosmic ray flux in interplanetary space, and so on. However, the exact relationship between them deserves deep investigation in the future.

\section*{Acknowledgements}

We thank the anonymous referee and the editor for their careful reading of our manuscript and for constructive comments and suggestions that improved the original version. The long-term data sets of the polar faculae applied in this work were taken from the Mitaka of the National Astronomical Observatory of Japan (http://solarwww.mtk.nao.ac.jp/en/db\_faculae.html), which provided the monthly time series of the polar faculae (recorded on the sunspot sketches) in the northern and southern hemispheres. The monthly time series of the sunspot numbers in the two hemispheres were freely taken from the website of the WDC-SILSO (World Data Center, Sunspot Index and Long-term Solar Observations), Royal Observatory of Belgium, Brussels (http://www.sidc.be/silso/datafiles). This work is supported by the National Key Research and Development Program of China (2018YFA0404603), the Joint Research Fund in Astronomy (Nos. U1631129, U1831204, U1931141) under cooperative agreement between the National Natural Science Foundation of China (NSFC) and Chinese Academy of Sciences (CAS), the National Natural Science Foundation of China (Nos. 11873089, 11903009, 11971421, 11961141001), the Youth Innovation Promotion Association CAS, the Yunnan Key Research and Development Program (2018IA054), the open research program of CAS Key Laboratory of Solar Activity (Nos. KLSA202002 and KLSA202016), and the major scientific research project of Guangdong regular institutions of higher learning (2017KZDXM062).

\bibliographystyle{mnras}
\bibliography{mnras} 

\bsp	
\label{lastpage}
\end{document}